\documentclass[journal]{vgtc}              

\onlineid{0}

\vgtccategory{Research}

\vgtcpapertype{Vis4Ed}

\title{Space to Teach: Content-Rich Canvases for Visually-Intensive Education}

\author{%
  \authororcid{Jesse Harden}{0009-0000-7827-1327},
  \authororcid{Nurit Kirshenbaum}{0000-0002-3587-1565}, 
  \authororcid{Roderick Tabalba}{0000-0002-6175-0353},
  Ryan Theriot,
  Michael Rogers,
  Mahdi Belcaid, \\
  Chris North,
  Luc Renambot,
  Lance Long,
  Andrew Johnson, and
  \authororcid{Jason Leigh}{0000-0001-7693-2814}
}

\authorfooter{
  \item
  	Jesse Harden is with Virginia Tech.
  	E-mail: jessemh@vt.edu.
  \item
  	Nurit Kirshenbaum is with University of Hawaii at Manoa.
  	E-mail: nuritk@hawaii.edu.
  \item 
    Roderick Tabalba is with University of Hawaii at Manoa.
    E-mail: tabalbar@hawaii.edu.
  \item 
    Ryan Theriot is with University of Hawaii at Manoa.
    E-mail: rtheriot@hawaii.edu.
  \item 
    Michael Rogers is with University of Hawaii at Manoa.
    E-mail: mlr2010@hawaii.edu.
  \item 
    Mahdi Belcaid is with University of Hawaii at Manoa.
    E-mail: mahdi@hawaii.edu.
  \item 
    Chris North is with Virginia Tech.
    E-mail: north@vt.edu.
  \item 
    Luc Renambot is with University of Illinois at Chicago.
    E-mail: renambot@uic.edu.
  \item 
    Lance Long is with University of Illinois at Chicago.
    E-mail: llong4@uic.edu.
  \item 
    Andrew Johnson is with University of Illinois at Chicago.
    E-mail: ajohnson@uic.edu.
  \item
    Jason Leigh is with University of Hawaii at Manoa.
    E-mail: leighj@hawaii.edu.
}

\abstract{%
With the decreasing cost of consumer display technologies making it easier for universities to have larger displays in classrooms, and the ubiquitous use of online tools such as collaborative whiteboards for remote learning during the COVID-19 pandemic, combining the two can be useful in higher education. This is especially true in visually intensive classes, such as data visualization courses, that can benefit from additional "space to teach," coined after the "space to think" sense-making idiom.  
In this paper, we reflect on our approach to using SAGE3, a collaborative whiteboard with advanced features, in higher education to teach visually intensive classes, provide examples of activities from our own visually-intensive courses, and present student feedback. We gather our observations into usage patterns for using content-rich canvases in education.
     
}


\teaser{
  \centering
  \includegraphics[width=.8\linewidth, alt={A view of a city with buildings peeking out of the clouds.}]{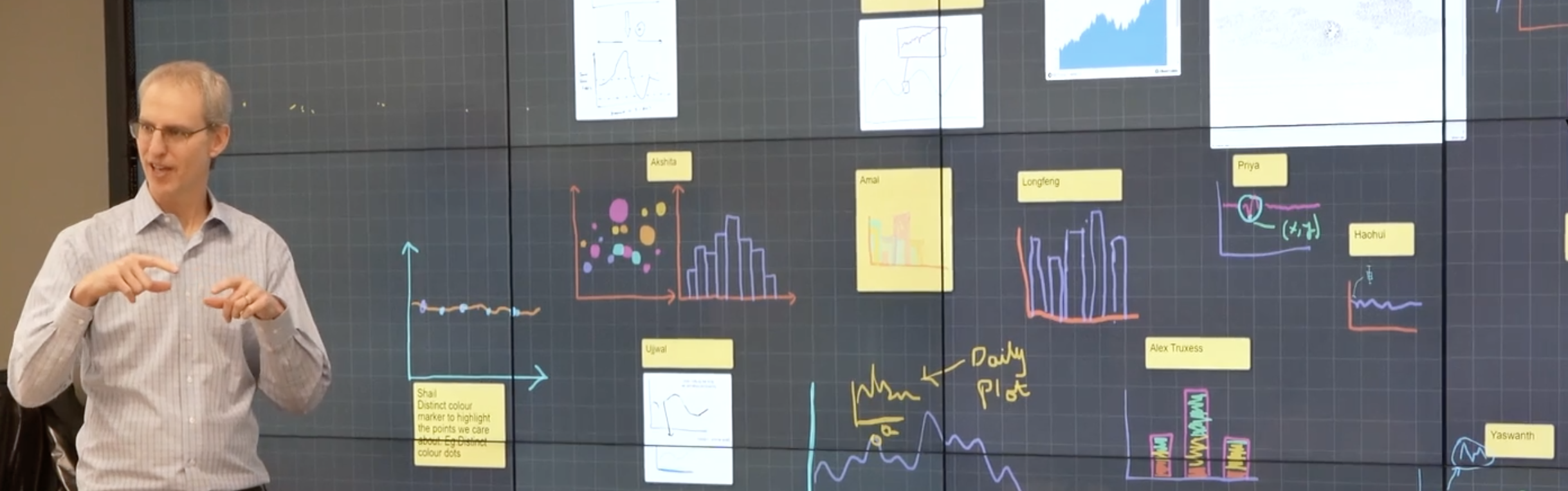}
  \caption{%
  	SAGE3 on a large display during a visualization class.
  }
  \label{fig:teaser}
}

\graphicspath{{figs/}{figures/}{pictures/}{images/}{./}} 

\usepackage{tabu}                      
\usepackage{booktabs}                  
\usepackage{lipsum}                    
\usepackage{mwe}                       

\usepackage{mathptmx}                  

\begin{document}


\firstsection{Introduction}

\maketitle

Content-rich canvases are virtual surfaces (often infinite) containing rich content arranged in spatially meaningful ways.
Typically implemented on digital whiteboards, content-rich canvases provide a platform for collaborative learning and engagement. These technologies held a special role during the COVID-19 pandemic, when educators of all levels searched for ways to augment remote learning. 
Indeed, researchers documented many success stories of enhanced student-led learning activities (e.g. \cite{10.1145/3513130.3558983,10.1145/3395245.3396433,ahmmad2021systematic,chan2023miro,kabil2023role,10343935}) using such tools, especially for online classes.
While learning has largely returned to in-person instruction, the use of shared digital canvases for co-located instruction sparks new opportunities, especially when combined with increasingly prevalent large displays. 
Indeed, we have long proposed using large displays for collaborative work and education \cite{10.1145/3395245.3396433}. 
With extensive experience in teaching higher education courses using SAGE3 \cite{tabalba_2023_10034793}, a content-rich canvas software we are part of the development team for, in classrooms with large displays, we identified patterns that portray how one can use content-rich canvases and large displays to augment education with their provided "Space to Teach", especially in classes that heavily depend on visuals and media, such as data science, game design, human computer interaction, and, of course, data visualization.


Visualization courses in higher education convey current theory in perception and data visualization, train students to critique and design visualizations, and cover practical aspects of programming frameworks for visualizations and/or relevant commercial tools \cite{kerren2008teaching}. 
With these broad topics in mind, many courses employ hands-on and collaborative activities in addition to traditional lectures. 
Visualization researchers and educators \cite{10310184} identified a need for further research pertaining to the educational methods that can efficiently help students to develop necessary skills for visualization work, as well as novel tools and environments that are beneficial for visualization education. 
We posit that using content-rich canvases like SAGE3 in classrooms with large displays (see Figure \ref{fig:teaser}) provides an edge for hands-on, collaborative course work like that in visualization and other visually-intensive courses.

Kirshenbaum et al. \cite{kirshenbaum2021traces} focused on content-rich canvases created via parallel content sharing during work meetings; they highlighted advantages of content-rich canvases in promoting \textbf{information continuity}, \textbf{information clustering} and \textbf{spatial memory}. 
This work explores such advantages in higher education. 
Many patterns we discuss here can be mimicked with other digital whiteboard tools (e.g. Miro) on large displays, although SAGE3 has some notable advantages which are discussed in the related works section.
We aim to show content-rich canvases work well with the nature of the material in visually-intensive courses like visualization, which usually relies on heavy use of images and interactive applications that showcase the variability and complexity of visualization examples. 
Our examples show content-rich canvases also support both instructor and student-led activities; both groups can freely share content, present, re-arrange, and sketch on such a canvas while engaging in educational activities; content
has permanence (information continuity), relevancy (information clustering) and spatiality (spatial memory) characteristic of content-rich canvases.

In this paper, we present six example patterns (Burst of Content, Side-by-Side, Content in Advance, Virtual Board Scroll, Bespoke Spaces, and Spatial Rearrangement), followed by student feedback evaluations and a discussion about the advantages, limitations, and future directions of using content-rich canvases and large displays in higher education. 


\section{Related Works}
To provide background for this work, we present some work relating to content-rich canvases and large displays, including benefits of large displays, online boards in education, and SAGE3. 
We then discuss challenges to visualization education and current efforts in that field.

\begin{figure*}[tb]
 \centering 
 \includegraphics[width=0.6\columnwidth]{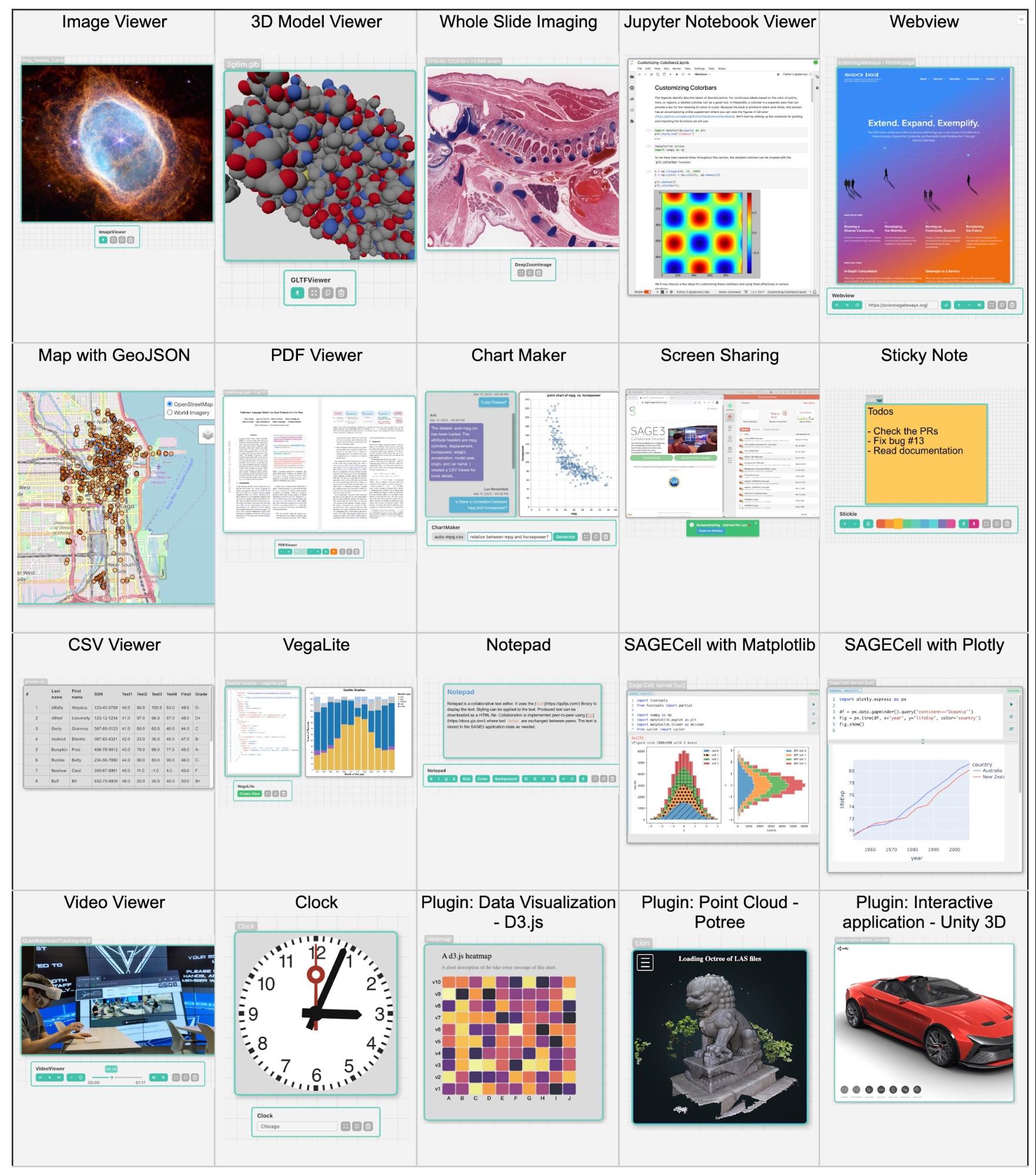}
  \includegraphics[width=1.2\columnwidth]{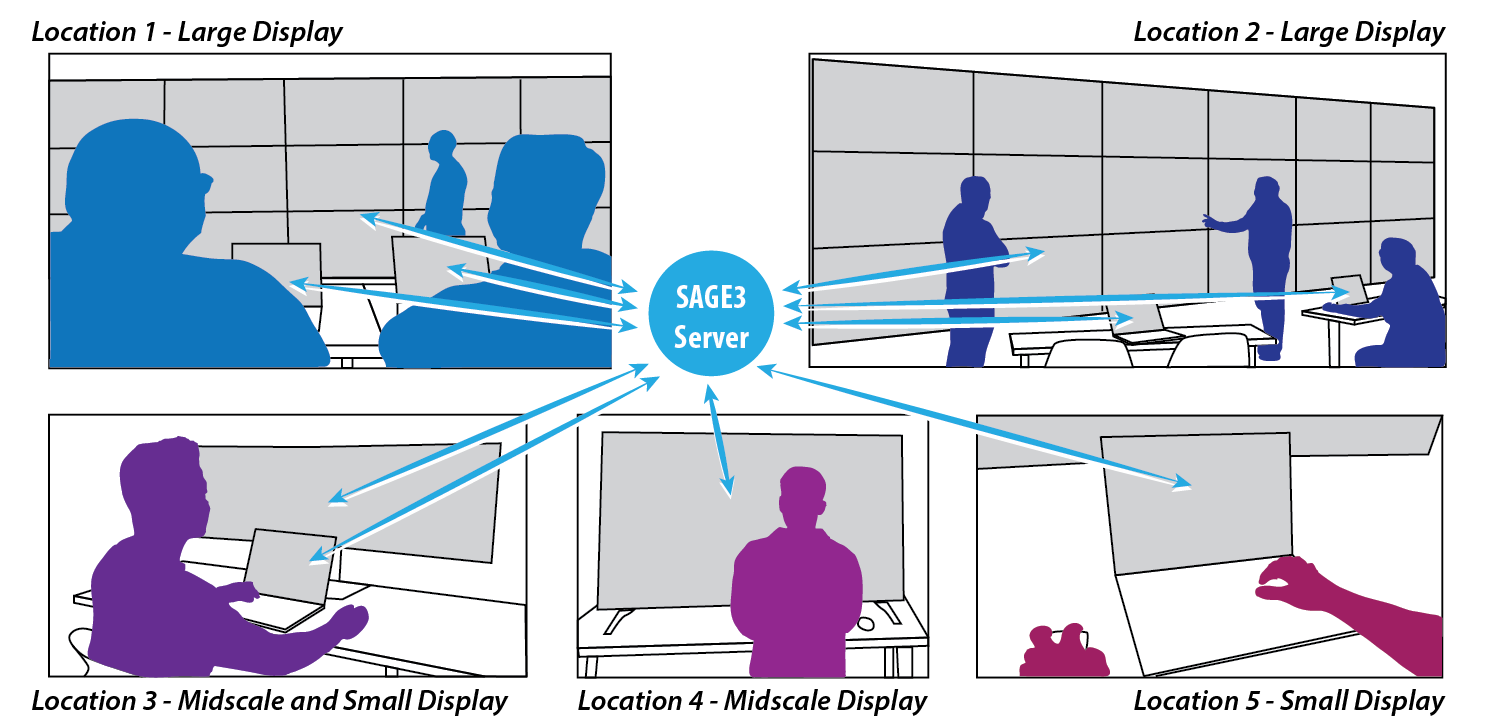}
 \caption{Left: A collage of apps that can be launched on a SAGE3 board. Right: A SAGE3 session can be shared across multiple locations: some locations may serve multiple users co-located in front of a large wall, while others may serve an individual with varied types of displays.}
 \label{fig:scenario}
\end{figure*}


\subsection{Large Displays \& Content-Rich Canvases}
The benefits of using large, high-resolution displays are well researched. 
They can positively influence spatial performance \cite{tan2003similar}, visualization and navigation tasks \cite{ball2005effects}, sense-making \cite{andrews2010space}, data analysis \cite{knudsen2012exploratory}, and daily work \cite{Bi:2009:CUL:1518701.1518855}. 
However, there are many challenges when it comes to controlling and working from a large display \cite{andrews2011information}.


Due in part to the COVID-19 pandemic, higher education saw a renaissance for online collaborative whiteboard platforms like Miro\footnote{https://miro.com/}, ConceptBoard\footnote{https://conceptboard.com/}, ExplainEverything\footnote{https://explaineverything.com/}, Google Jamboard\footnote{https://jamboard.google.com/}, Sketchboard\footnote{https://sketchboard.io/} and more \cite{ahmmad2021systematic}. Many publications describe experiences and study results from this period (e.g. \cite{chan2023miro,kabil2023role}). These works show that using whiteboard platforms encourage collaboration and creativity, but have a learning curve. 
Modern whiteboard software supports file sharing (e.g. images and PDF files), use of colorful notes similar to physical Post-it notes, drawing features, and more. More features increases the richness of the resulting boards, hence our term of content-rich canvases; at the same time, tool mastery becomes more difficult.

While some patterns described in this paper can be implemented on other online whiteboards, we emphasise SAGE3 since it differs from other whiteboard platforms in three significant ways: it was designed with large displays and varying display sizes in mind as seen in Figure~\ref{fig:scenario}, it is open-source software developed as an NSF funded research project and thus is, unlike commercial products, open to modification, and it has a backend designed to support computational notebook cells, a feature that was not used during the educational activities reported in this paper, but is likely to be prominent in future use.
Finally, SAGE3 features many other applications, as seen in Figure~\ref{fig:scenario}, including a WebView app that enables internet browsing within a board, PDF viewer that can be extended to view multiple pages side by side, and screenshare (simultaneous multiple screenshares are supported). 
More information about SAGE3 is available in SAGE3's public repository \cite{sage3github}.

\subsection{Visualization Education}
Teaching and learning the skills of creating and analysing visualization spans a variety of topics, from designing visualization education tools for young children \cite{8807271,9222251} to evaluating the relevance of existing tools to visualization courses \cite{9453917,6295799}. For higher education, some challenges were verbalized by the ACM SIGGRAPH Education Committee in 2013 \cite{6562722}. Educators at that time noted that:
\begin{itemize}
    \item  there is an increase in nontechnical students learning visualization due to its relevancy to the modern workforce,
    \item Hardware technology and visualization algorithms have progressed and may affect computer science students taking visualization courses,
    \item visual analytics are increasingly used in practice, leading to a strong need for highly interactive visualization,
    \item  user-centered design and evaluation of visualizations is necessary.
\end{itemize}

More recently, Bach et al. articulated challenges for visualization education\cite{10310184}. They arrange these challenges around the themes: people, goals and assessments, motivation, methods, environment, materials, and change. Largely, many of the issues articulated in 2013 remain and are even amplified; A boom in data science education leads to a very varied body of students interested in visualization, the class settings have developed beyond the simple in-person lectures to online and hybrid classes, and the technological advancements lead to more complex visualization systems to learn, design, and most importantly, adapt to. Bach et al. raised 43 specific research questions based on the then-current state of visualization education.
We would like to highlight the following questions from Bach et al.:
\begin{itemize}
    \item In their discussion on Methods and the need to foster core skills around visual representation and interaction, the authors ask (Q21-Q23) for ways to develop such skills, and doing so while leveraging play  and/or using sophisticated novel tools.
    \item In their discussion on Environment and the need for providing environments for hands-on, creative,
and collaborative work, the authors ask (Q27, Q28) about affect and affordance of specific environments that can support data visualization education.
\end{itemize}
We propose content-rich canvases like SAGE3 are a tool that can be used to answer these questions and the patterns described below provide opportunities for non-conventional activities appropriate for visually intensive material due to the affordance of space unique to an environment using content-rich canvases and large displays.

Of course, there are many outside-the-box educational explorations in visualization. Beasley et al. \cite{beasley2020leveraging} investigate how integrating peer feedback throughout the semester can improve students' engagement in visualization classes.
Boucher et al. \cite{10344064} identify the potential of using educational data comics, discussing how it supports visualization activities while appealing to a variety of audiences. Boucher and Adar \cite{he2016v} describe their workshops to help students through the visualization design process with inspiration, layout, and domain specific cards. Adar and Lee-Robbins \cite{adar2022roboviz} devised a  framework around a visualization class' final project that overcomes some issues of traditional project assignments like the need to clean data and the difficulty of evaluating project outcomes by using an engaging vehicle of a game called Roboviz. There is a continued need for innovation in visualization education, and this paper helps fill this niche.

\section{Content-Rich Canvas Usage Patterns in Education}

\begin{figure}[tb]
 \centering 
 \includegraphics[width=1\columnwidth]{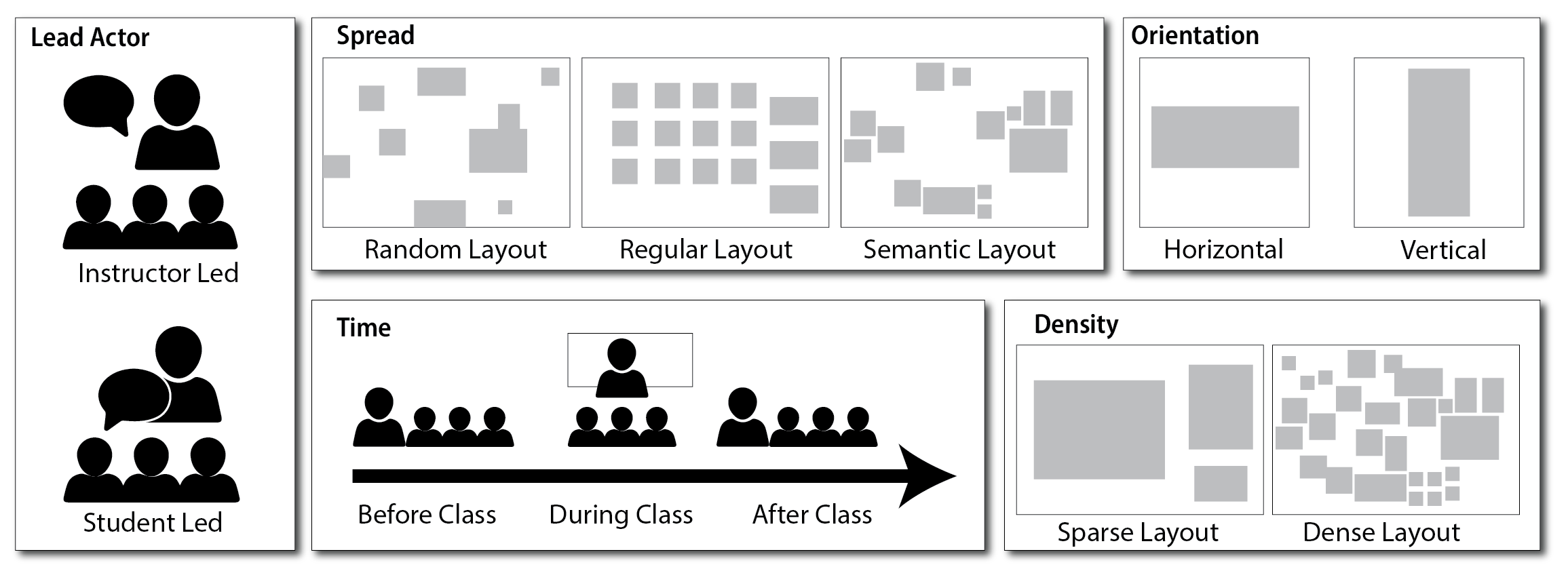}
 \caption{Elements of content-rich canvas usage patterns in education.}
 \label{fig:elements}
\end{figure}

Classes (in higher education and otherwise) traditionally revolve around a single instructor-controlled view, which often consists of a slide deck presentation or the instructor screen-sharing their personal view. When students are asked to take part in the presentation, a ritualistic "passing of the cord" is performed from student to student as each gets access to the communal view (usually, by plugging their computer into a display) on their turn. This conventional, sequential, single-view mode of content sharing can be improved upon with the space provided by a content-rich canvas and large display setup.
 
While teaching a variety of courses on topics such as visualization, VR, data science, and game design, members of our team have periodically written notes regarding how they use SAGE3 in their classrooms.
These notes were analyzed by the authors using a grounded theory approach to identify several patterns and strategies of usage in the classroom. 
These patterns are not mutually exclusive, and the subset of patterns used in any specific course are influenced by the course topic and the instructor’s personal style of teaching.

\subsection{Elements of the Patterns}
As mentioned in prior work \cite{kirshenbaum2021traces}, three properties of content-rich canvases are \textbf{permanence}, or information continuity, \textbf{relevancy}, or information clustering, and \textbf{spatiality}, or spatial memory.

Below, we describe the base elements of \textbf{orientation}, \textbf{spread}, \textbf{density}, and \textbf{time}, which play a role in our patterns. We also add the element of \textbf{lead actor}, which nods to the education-specific hierarchy of instructor and students. Figure \ref{fig:elements} summarizes these elements.

\paragraph{Orientation}
When creating a content-rich canvas, the placement of content may result inn creating a mostly horizontal layout or a mostly vertical layout. Modern displays are usually more wide than they are tall, so the horizontal layout is prominent. Some content could be mostly vertical in nature, such as data tables or linear computational notebooks. Vertical layouts may play another role - one that indicates the passage of time; see the \textit{virtual board scroll} pattern below. 
\paragraph{Spread}
Multiple units of content (e.g. apps in SAGE3) can be spread across the space in a random, regular, or semantic layout. The random layout is characteristic of multiple collaborators adding content (either by uploading files or starting application on the board) at approximately the same time without the help of \textit{bespoke spaces} (see pattern below). Regular layouts are usually started by one user trying to establish a specific structure for an activity. Semantic layouts are formed when users move content around the board to indicate relationships between units of content (see the \textit{spatial rearrangement} pattern).
\paragraph{Density}
Layouts can be sparse or dense. Sparse boards can contain as little as a single piece of content (for example, a screen-share or a PDF of a deck of slides) or some other small number (for example, a PDF with the syllabus and a stickie with the assignments schedule). The dense boards have many more units of content on them; this can be useful for many activities, but can also be overwhelming and "busy" at times. The \textit{side-by-side} pattern usually creates a sparse layout and the \textit{burst of content} pattern is likely to be on the dense side.
The advantage of an infinite board with the large display demarcating the current area in use is that a board can be sparse and dense in different areas of the canvas and by using navigation the instructor can control the extent of the content in focus. This is used extensively classes (see the \textit{virtual board scroll} pattern).
\paragraph{Time}
The element of time is relevant in a content-rich canvas where we can expect time to affect the use of space (see "Traces of time through space" \cite{kirshenbaum2021traces}). In the educational setting, we consider the key time points of before, during, and after class. Intuitively, the period of time during class is when the board is under heavy use, but since the content persists there is great value with preparing \textit{content in advance}, which is one of the design patterns below, and material can be downloaded from a board or reviewed as needed after the class. 
\paragraph{Lead Actor}
Though we are not drawing a line between the instructor using the board and the students using it, educational activities are usually led by an instructor (i.e. lecture, demonstrating software, showing and critiquing examples, etc.) or by one or more students (i.e. project presentation, brainstorming in groups, solving problems in class etc.) The extent of freedom given to students during class is dependant on the instructor and their chosen activities. 

\subsection{Patterns}
In this section we describe recurring patterns of use that we came across while teaching visually intensive classes using the content-rich canvas software SAGE3.

\subsubsection{Burst of Content}
We frequently find SAGE3 useful for brainstorming activities; 
for example, students in a video game design class were asked to break into small groups, with each group creating a board within the classroom's room that was used to discuss project ideas based on given criteria.
Students could use their board how they wished, organizing reference documents and stickie notes. 
The instructor visited the board of each group and provided feedback. 
The students could share their board on the large display when asked to present their ideas. 
This is a common form of activity seen in classes that require group-work and employ a design process. 
In terms of the elements discussed above, this would be a student-led content-rich canvas that is created either during or after class, and is likely to take the form of a dense, semantic layout.

However, this pattern goes beyond group brainstorming. 
The \textbf{Burst of Content} pattern indicates that many sources of content are shared at the same time; 
this can be useful when the instructor is posing a question or a task to all the students, and all the students are expected to respond more or less simultaneously. 
This would usually create a content-rich canvas that is initially led by the instructor with a random, dense layout formed as students post their responses, and later turned to a canvas with semantic layout either at the hands of the instructor or the students. 
Overall, this pattern works well for ideation, group discussion, simulation/play, and compare/contrast activities.

\subsubsection{Side-by-Side}
Often, teaching materials are incremental or referential: 
the slide about drawing a graph is built on the content of the slide containing an adjacency matrix, a slide about techniques for zooming relies on the definition of viewports a few slides back, or an instructor would like to compare the code for map() and reduce() which appear on different slides. 
Normally, instructors find themselves duplicating information between slides or flipping back and forth; 
a content-rich canvas with a large display eliminates this need.

The advantage of content-rich canvases like SAGE3 is collaborative side-by-side placement of teaching materials: 
a video next to an equation it refers to, live screenshare next to step-by-step instructions, or multiple slides side-by-side for easier referral.
This latter feature is enabled by SAGE3’s built-in PDF viewing app, which supports multi-page view and the physical navigation afforded by the large display.
Some instructors use a spread of as many as 4 consecutive slides from their slide deck. 
Any app can also be duplicated when the instructor wants to show side-by-side different elements from the same source. 

This pattern can be used by an individual, be it an instructor or a student, or by multiple students working together. It naturally leads to horizontal layouts and often takes a sparse form. This pattern works well for narratives (i.e. presenting material in a linear order) and for compare/contrast activities.

\subsubsection{Content in Advance}
One author used to create a multi-tabbed browser window with each tab directed at an example to be explored during class. 
This method of teaching felt awkward; 
despite preparing in advance, they had to search for the "correct" tab when they reach the relevant material, and switching tabs would put away the lecture slides, making it harder to relate lecture content to the example. 
This issue can be solved by preparing content in advance on a content-rich canvas.

This pattern supports a flexible sequence for presentation of content, and 
a presenter can use this pattern in the following way: 
before the presentation starts, the presenter loads the board with content, such as slides, webviews, and videos.
While presenting, the presenter highlights content as needed. 
Spreading materials in this way gives students access to more modalities of the material; 
students can easily download files from the board, be it PDF files, images or video files.  
With the SAGE3 webview, the presenter can quickly switch between talking about an example to interacting with an example in a live webview within the board. 
Students can follow the interactive examples on the display or use the webview to launch the web pages in their own browsers.

The main element relevant to this pattern is time:
there are no limitations on the kind of layout the presenter creates, and the presenter can be either the instructor or a student. This pattern works well for narratives (linear or non-linear) and for compare/contrast activities.

\subsubsection{Virtual Board Scroll}
Many large classrooms contain vertically sliding blackboards, where instructors write their notes on a "fresh" board and slide it up when the board was filled revealing another layer of sliding board. 
When that one was filled, up it goes, and the previous board was brought down, cleaned, and used again for notes. 
This was a physical way to maintain persistence of the notes, at least for a while, giving students more time with them. 
The instructor didn't need to destroy old content because they could "scroll" in a way that gave them a new space to fill with new content, which is the essence of this pattern. 
While all content remains on the board, and students can access them as they wish, the instructor sets the large display's viewport into the board, virtually scrolling the material to their current focus.
This pattern is mostly instructor led (unless a student is presenting), occurs during class, and can use horizontal and/or vertical space, if that verticality is divided into multiple semantically relevant horizontal displays the presenter would like to scroll through.  

\begin{figure}[tb]
 \centering 
 \begin{subfigure}[b]{0.7\columnwidth}
  \includegraphics[width=\textwidth]{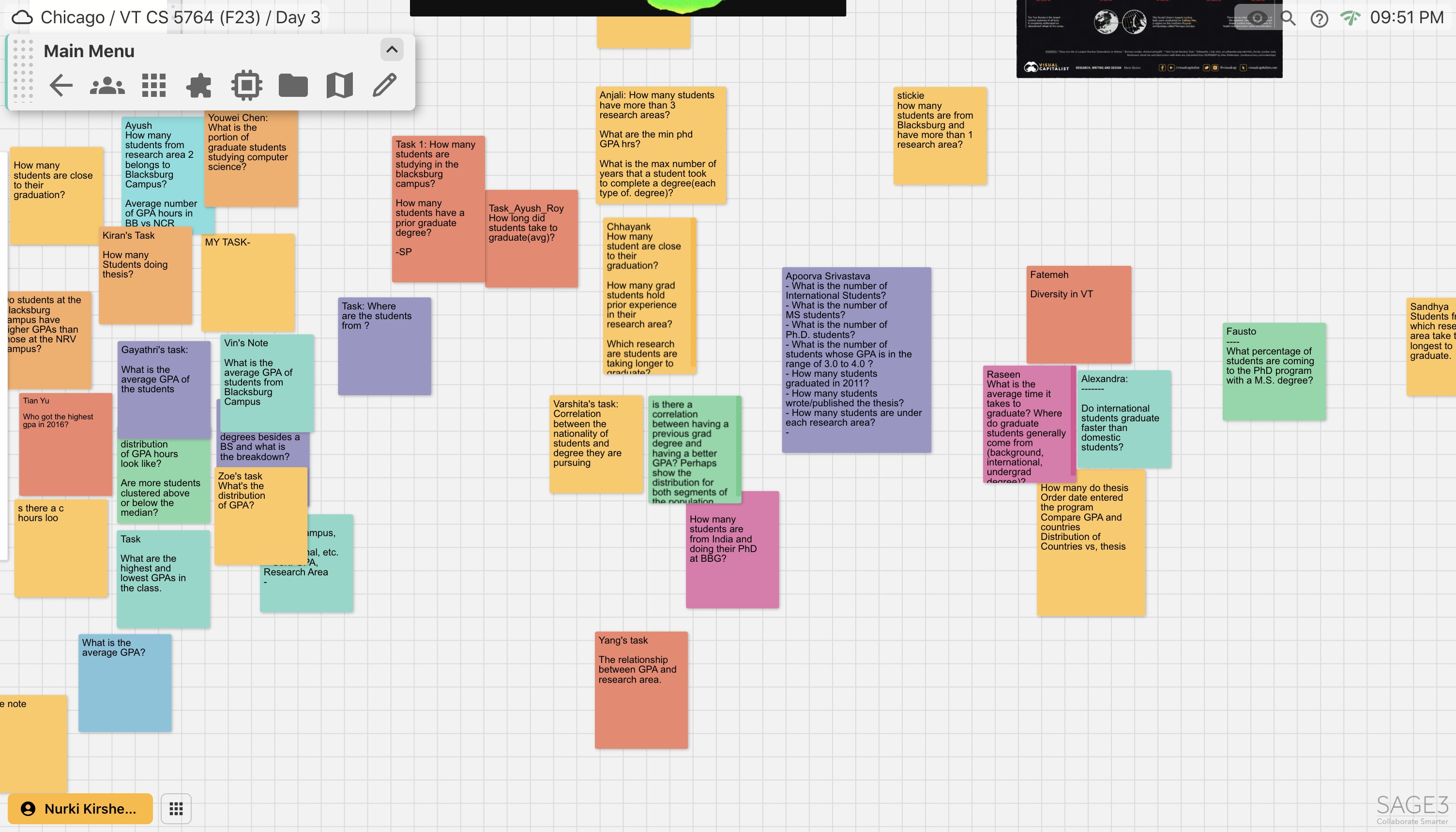}
  \caption{Brainstorming board.}
  \label{fig:case1}
  \end{subfigure}
  \begin{subfigure}[b]{0.7\columnwidth}
  \includegraphics[width=\textwidth]{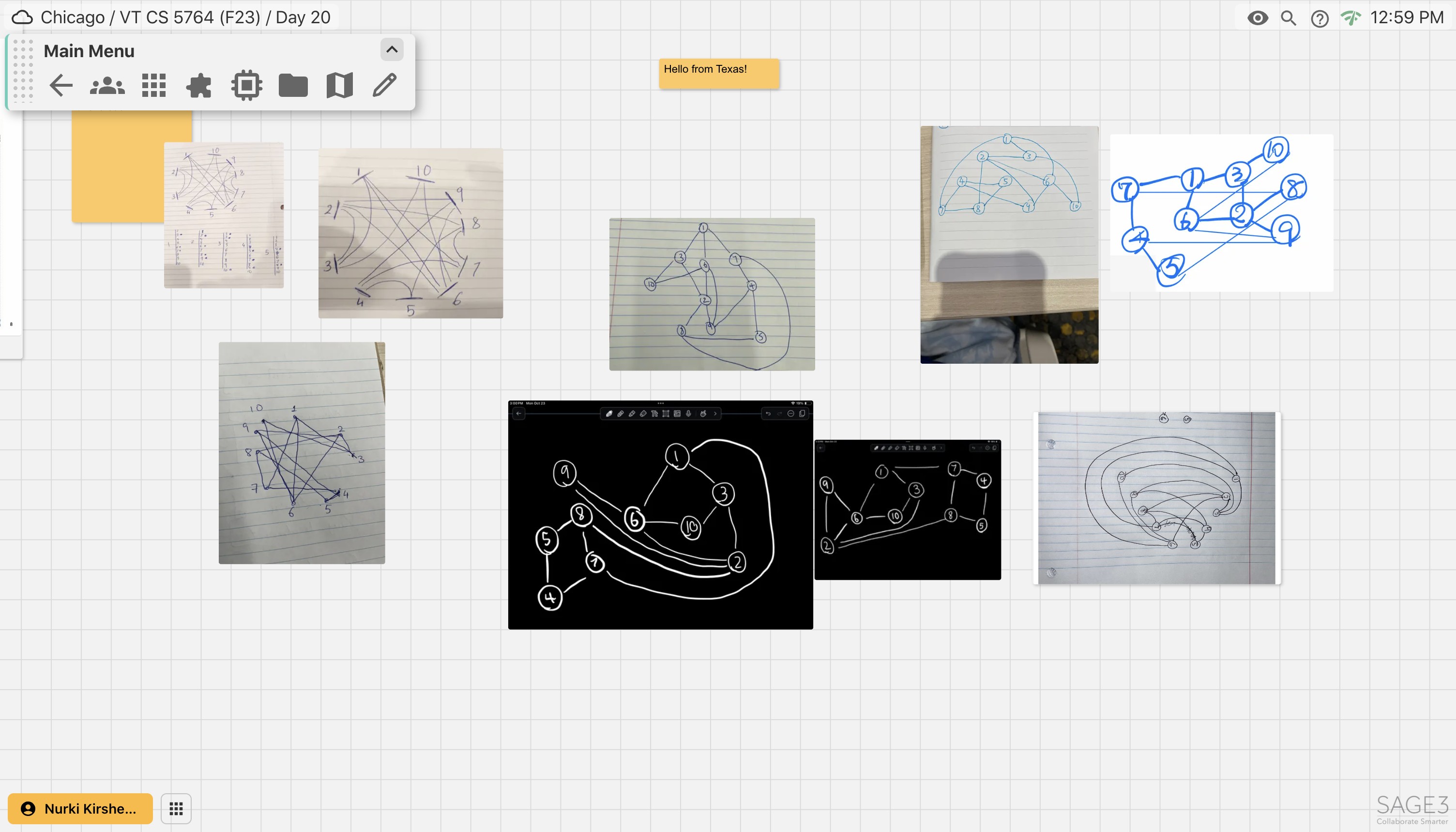}
  \caption{Clustering graphs.}
  \label{fig:case2}
  \end{subfigure}
  \begin{subfigure}[b]{0.7\columnwidth}
  \includegraphics[width=\textwidth]{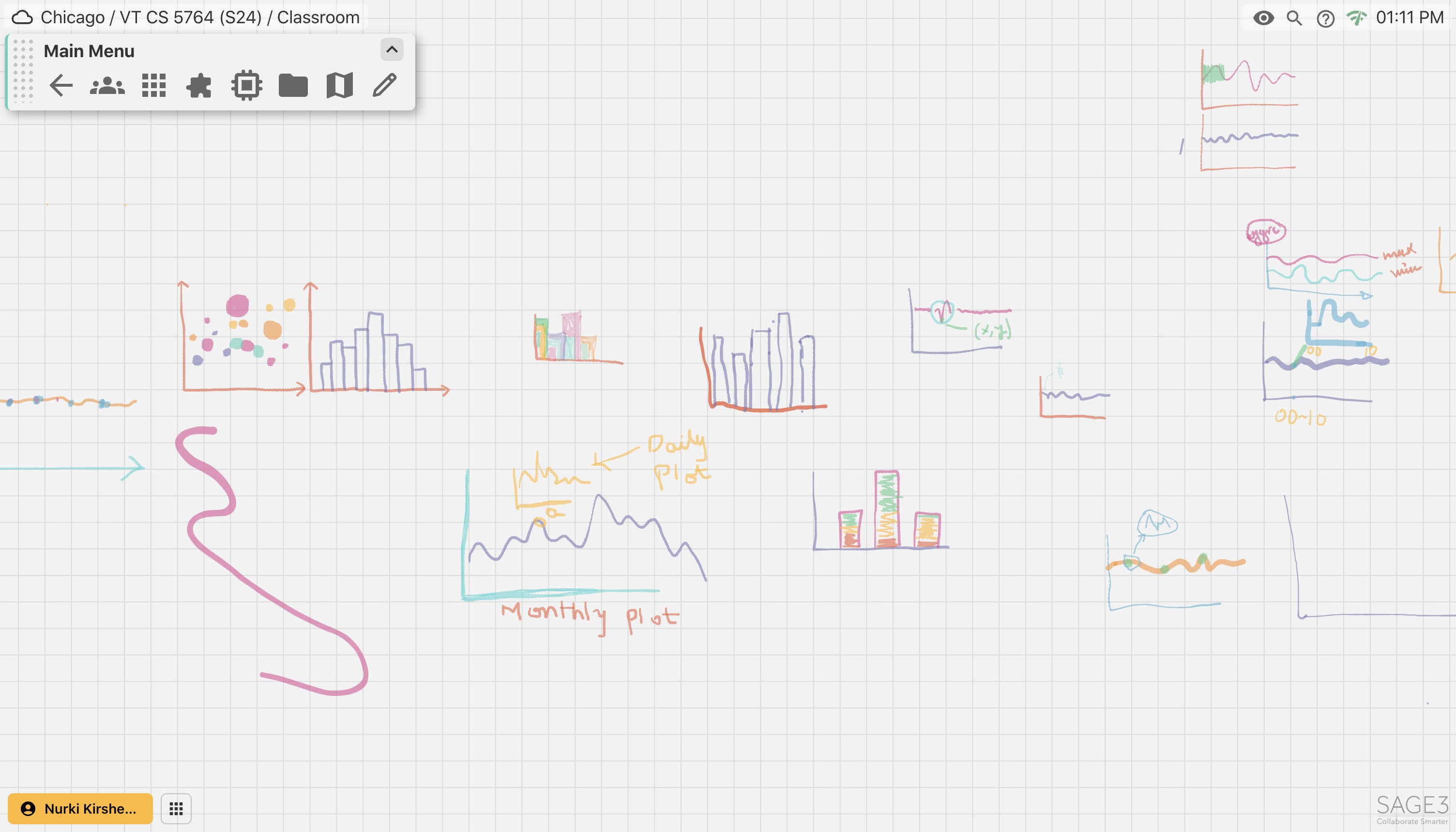}
  \caption{Visualization sketched in annotations.}
  \label{fig:case3}
  \end{subfigure}
  \begin{subfigure}[b]{0.7\columnwidth}
  \includegraphics[width=\textwidth]{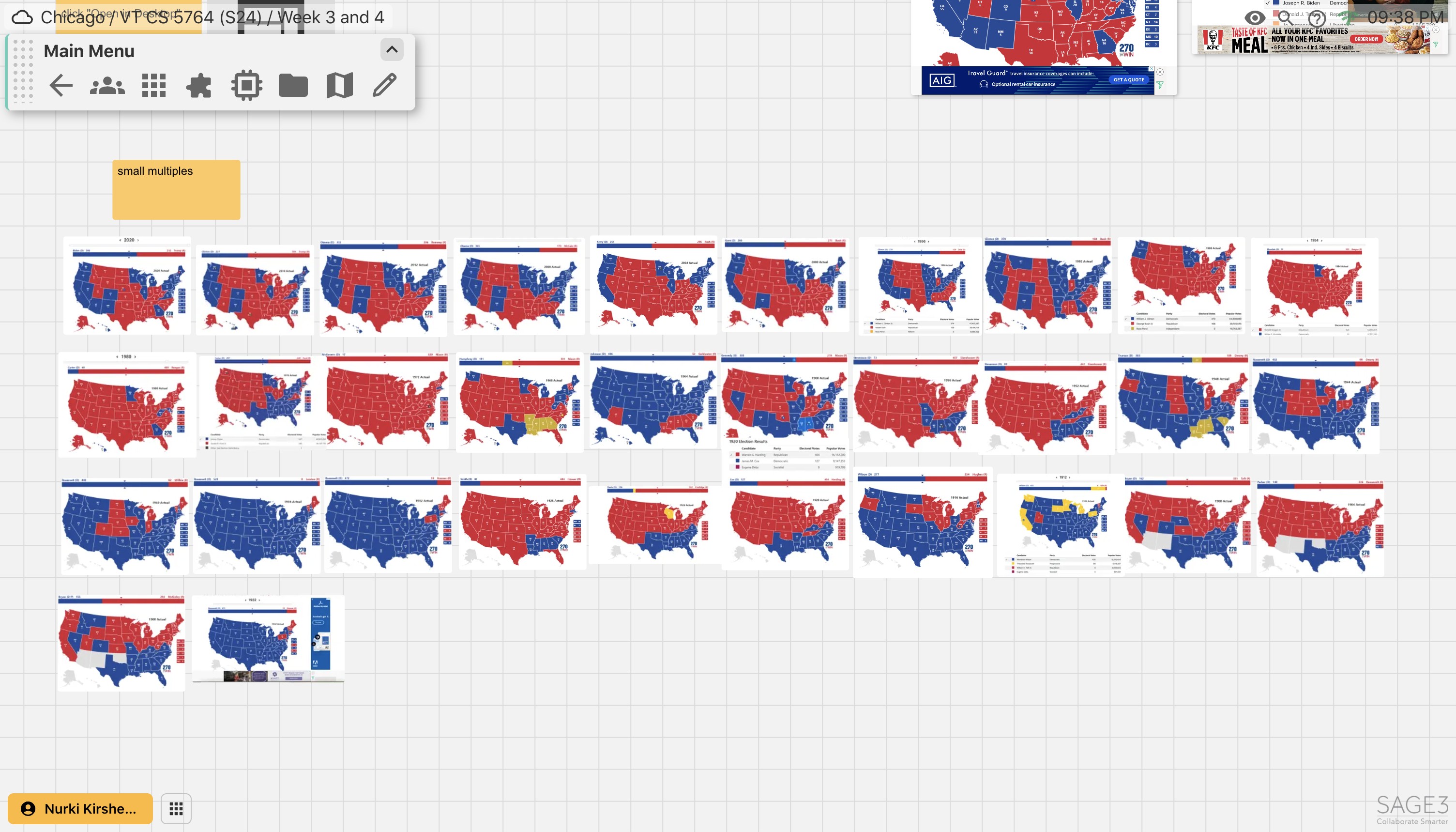}
  \caption{Small multiples created collaboratively.}
  \label{fig:case4}
  \end{subfigure}
 \caption{Example boards demonstrating burst of content. }
 \label{fig:education1}
\end{figure}

\subsubsection{Bespoke Spaces}
When facing an infinite canvas, users need a starting point; 
when using a large display, users may not know how to position their content, assuming that content does not cover the full screen, so it makes sense to divide space into areas dedicated to specific uses. 
In addition to showing their slide deck on the large display, instructors may have a part of the visible viewport show the agenda for the class, and another part for student questions.
This approach relates to general trends of using abundant display real-estate, like how PC users with multiple displays tend to
assign displays roles like "reference" and "active work").

Another way to use bespoke spaces is to have a section of the board for students to post content;
this can be used with regularly placed and labelled stickie notes so every student knows exactly where on the board to post. 
Students can also create and use their own boards, as mentioned in the brainstorming scenario above, forming an organizational separation from all other material rather than a spatial separation.

Using regular or semantic layout, this pattern is useful for students and instructors any time during or outside of class, and is appropriate for narrative, ideation, simulation/play, and compare/contrast activities.

\subsubsection{Spatial Rearrangement}
We keep going back to the importance of using the space, looking at different orientations, densities, and meaningful or random ways to arrange the spread of content on the board. 
This pattern draws attention to the act of moving content around, which can be done by both instructor and students at any time. 
Often, rearranging content is done to form a specific spread, such as changing a random spread into a semantic or regular one. 
Rearrangement can also be used to denote relevance or irrelevance of specific content;
when content is irrelevant to the current discussion, the app window containing it can be closed or be moved away from the area in focus, out of sight but not necessarily out of mind, where it remains a referent available for access when needed. 
Rearrangement may also be needed to transition from one orientation to another, such as when the display being used as a communal wall is changed or the board's window is resized.

This pattern is the bread and butter of any extensive use of content-rich canvases and can be used in any activity.

\subsection{Examples}
In this section we share some examples of activities and their accompanying boards taken directly from a visualization class delivered by one of the authors;
the examples were brought up during an interview with them (see further detail in the Section 5.2.2).

Brainstorming is very commonly performed on large displays, with one such activity given in the visualization class:
Given a dataset, students were asked to come up with thoughtful questions they could answer with visualizations. After the questions were posted, the instructor and student clustered the questions based on similarity to enable exploration and analysis of the design space for visualizations of said dataset. The resulting board is shown in Figure \ref{fig:case1}. This is a simple use-case for the \textit{burst of content} and \textit{spatial rearrangement} patterns.
 
 Another activity using these patterns was given during another class;
 the students had to sketch a graph based on a provided adjacency matrix, a common task in visualization classes, such as shown by Kerren et al. \cite{kerren2008teaching}. 
 The sketches were uploaded to the board and the instructor and students clustered them based on various characteristics such as graph layout. 
 The resulting board is shown in Figure \ref{fig:case2}.
 
 Other examples of the \textit{burst of content} pattern are shown in Figures \ref{fig:case3} and \ref{fig:case4}. 
 In the former, students are given a dataset, explore it and sketch (using the annotation tool) what kind of chart they would create for it. 
 The latter shows a more complex exercise:
 creating a collaborative multi-view visualization in which students had to post visualization around a theme, US election outcomes, and organize them according to a dimension, the election year. 
 Together they created and analysed a small multiples example of visualization.

\begin{figure}[tb]
 \centering 
 \begin{subfigure}[b]{0.7\columnwidth}
  \includegraphics[width=\textwidth]{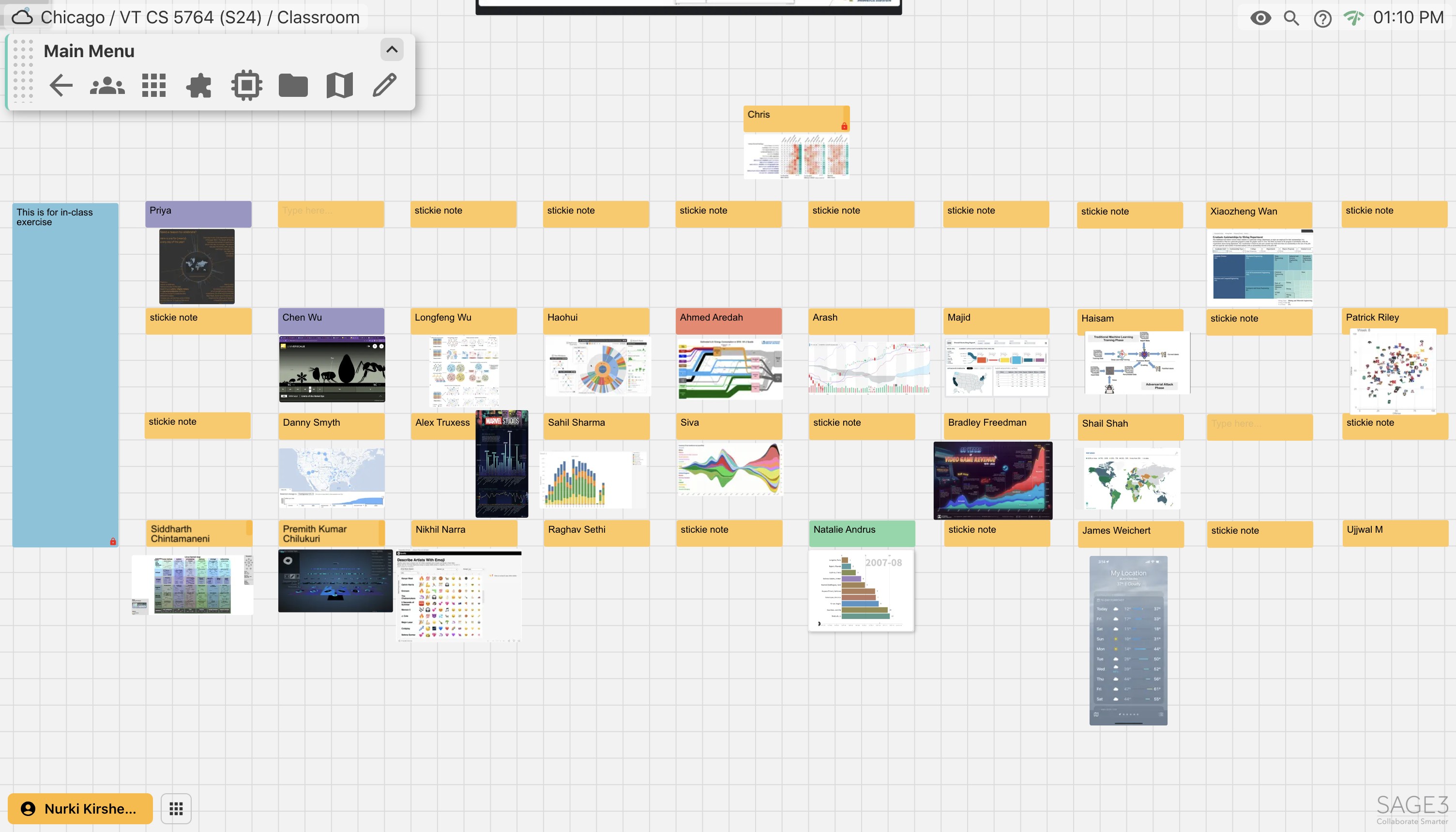}
  \caption{Post a visualization in a bespoke space.}
  \label{fig:case5}
  \end{subfigure}
 \begin{subfigure}[b]{0.7\columnwidth}
  \includegraphics[width=\textwidth]{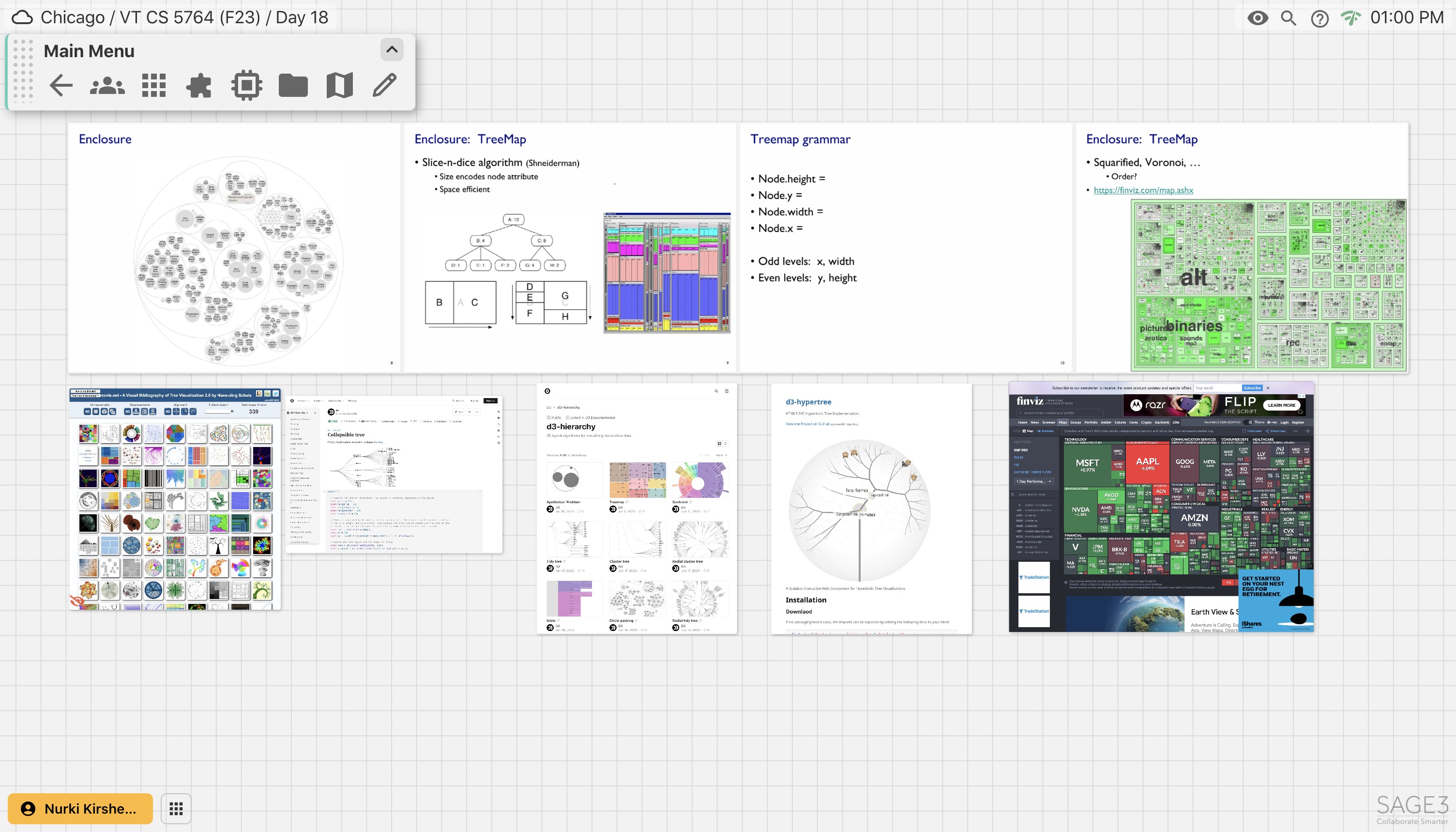}
  \caption{Side by side and in advance, posting interactive examples for the lecture.}
  \label{fig:case6}
  \end{subfigure}
  \begin{subfigure}[b]{0.7\columnwidth}
  \includegraphics[width=\textwidth]{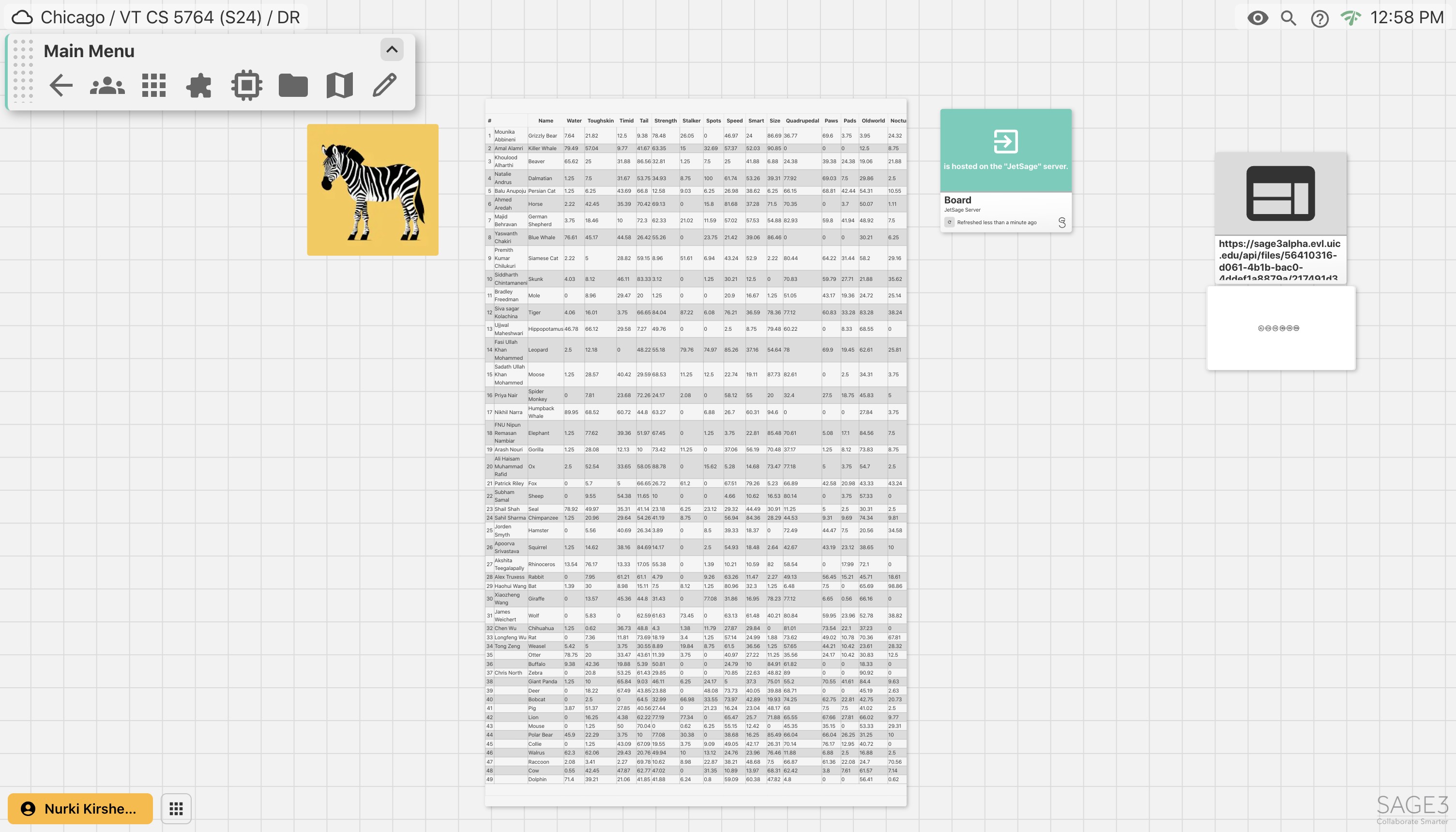}
  \caption{Dataset is presented.}
  \label{fig:case7}
  \end{subfigure}
  \begin{subfigure}[b]{0.7\columnwidth}
  \includegraphics[width=\textwidth]{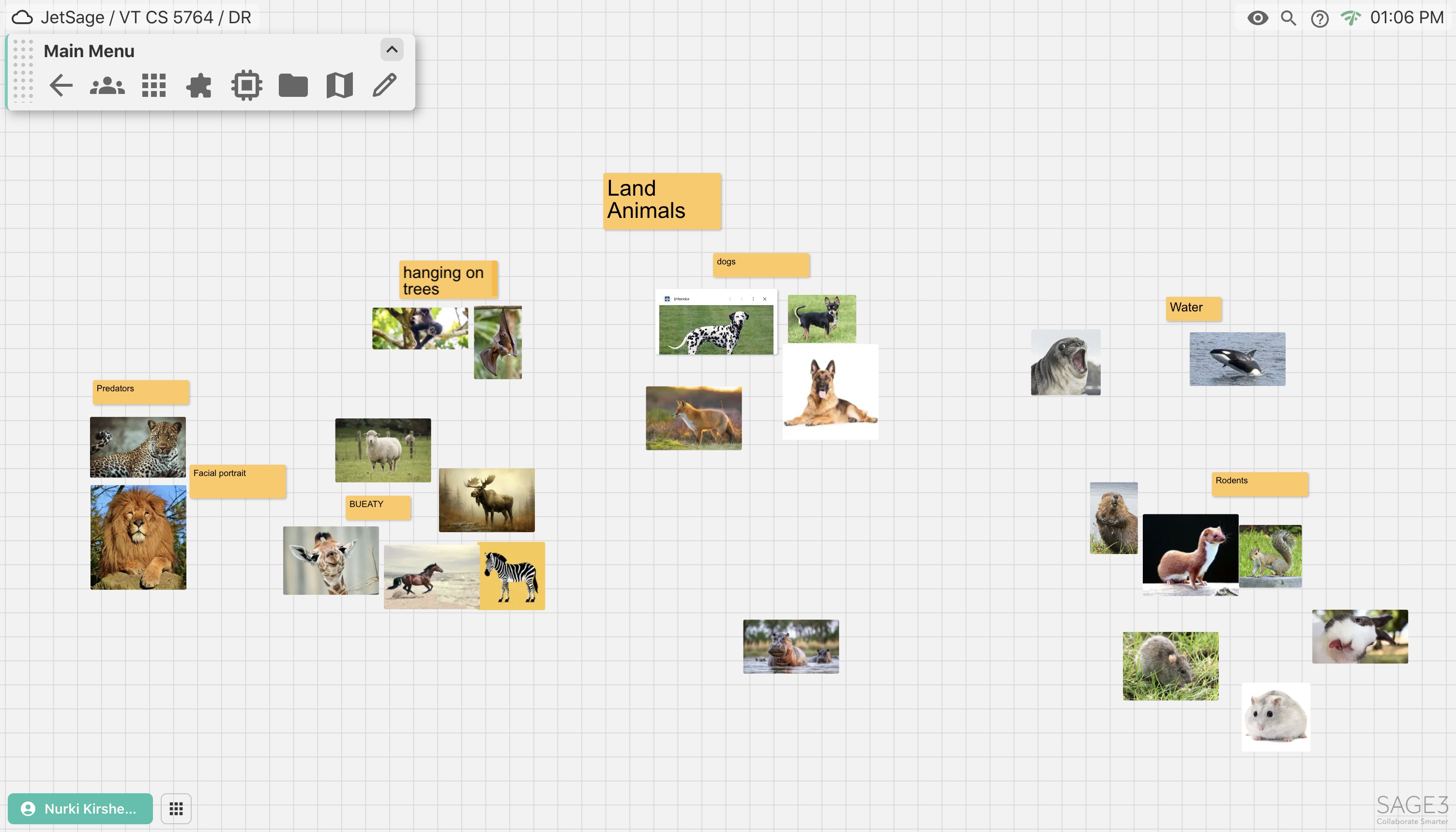}
  \caption{MDS simulation exercise, where would you place your record in 2D considering the multi-dimensional similarity.}
  \label{fig:case8}
  \end{subfigure}
 \caption{Example boards demonstrating various usage patterns,}
 \label{fig:education2}
\end{figure}

An example of simulation activity that uses \textit{burst of content} and \textit{spatial rearrangement} can be seen in Figures \ref{fig:case7} and \ref{fig:case8}. 
During a class on Multi-dimentional Scaling (MDS), students received a dataset on animals for their review, and each student was assigned an animal from the data; 
on a new board, they perform an activity simulating MDS. 
Students place an image of their animal on the board, and then move their image closer to similar images and further away from dissimilar images, encouraging them to wrestle with the relationship between distances and similarity in two dimensions instead of higher dimensions.

One of the interviewee's favorite methods uses the \textit{bespoke spaces} pattern effectively.
With an ordered array of stickies on the board, they assign students a safe space to work and track their class participation, such as shown in  \ref{fig:case5}. 
During the first class, the instructor arranged stickie notes on a SAGE3 board to mimic the seats in class and asked students to write their name on the stickie corresponding to their seat;
later, students found a visualization of interest and shared it with the class by placing their found image under their stickie.



An example of a class that is more lecture-centric is shown in Figure \ref{fig:case6}, which is a good example of the \textit{side-by-side} pattern along with \textit{content in advance}. 
In this class the instructor focused on TreeMaps. Before class, the instructor prepared examples in the form of interactive apps that demonstrate their slide-deck material, and during class they would seamlessly transition to these interactive examples.

These examples show glimpses of other content, meaning \textit{virtual board scroll} was used to isolate the activity's area on the board.



\section{User Feedback}
While our usage patterns are the main contribution in this paper, we would be remiss to not include students' feedback for SAGE3 and SAGE3-based classes. This section covers an evaluation survey administered in two visually intensive classes at the University of Hawaii and another evaluation survey administered in two visualization classes at Virginia Tech. This latter evaluation was followed by an interview session, where two of the authors interviewed the instructor, who is also an author of this paper.

\subsection{Evaluation 1}
In Spring 2023, we distributed a survey among students of two classes that used SAGE3 for in-class instruction; 
both involved game design projects in small groups. 
One class was a Junior level class called "Computational Media System" and the other was a Sophomore level class called "Video Game Design and Development."
Both classes met in-person on the University of Hawaii campus, and were taught by different instructors from the SAGE3 group. 
Both instructors used SAGE3 as the substrate for delivering lectures and student activities, although the university wide learning management system, Laulima, was also used for content delivery and assignment submissions.

Surveys for evaluating and improving SAGE3 were brought up in class near the end of the semester, and it was clearly explained that submission is anonymous and no repercussions for not submitting it or expressing negative views on SAGE3 would occur. 
26 students from the computational media class and 23 students from the game design class submitted the survey, making a total of 49 respondents.

The survey had two sections: 
the first gauged their experience with various SAGE3 features, and the second prompted attitudinal responses regarding the use of SAGE3.
Following the Likert scale questions of the survey, students could write comments elaborating on their views.

\subsubsection{Results}
\begin{figure}[tb]
 \centering 
  \includegraphics[width=\columnwidth]{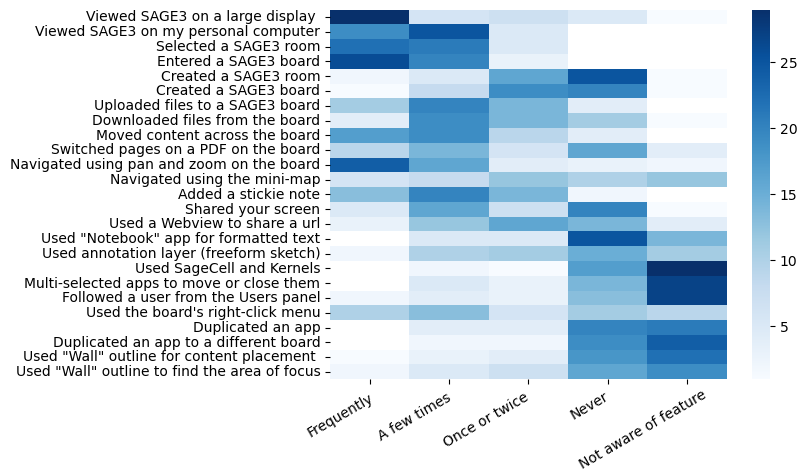}
 \caption{Heatmap showing how the students used different features of SAGE3. }
 \label{fig:experience}
\end{figure}

\begin{figure}[tb]
 \centering 
  \includegraphics[width=\columnwidth]{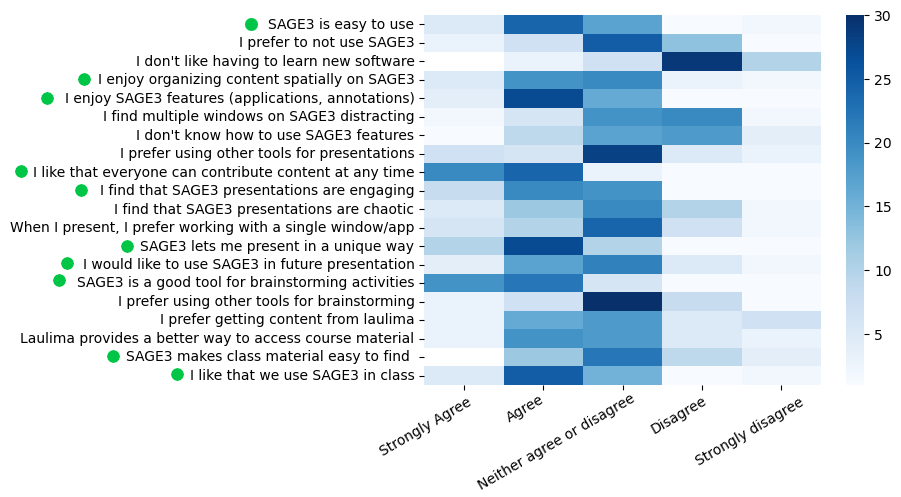}
 \caption{Heatmap showing students' attitude towards using SAGE3. Questions marked with green indicate positive attitudes. }
 \label{fig:attitude}
\end{figure}

The survey responses for the first section on experience with features are summarised in Figure \ref{fig:experience}, and the responses for the second section on their attitude towards SAGE3 are summarised in Figure \ref{fig:attitude}.

We see a higher concentration of feature use for basic features like entering or navigating a board;
for more advanced features, like annotation, SAGECells, follow feature, and app duplication, students admitted to not using them, and were often unaware of them.

\begin{figure}
    \centering
    \includegraphics[width=.85\columnwidth]{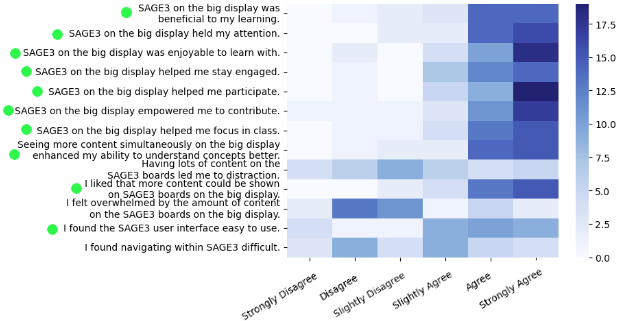}
    \caption{Student attitude towards the use of SAGE3.}
    \label{fig:chart3}
\end{figure}

\begin{figure}
    \centering
    \includegraphics[width=.85\columnwidth]{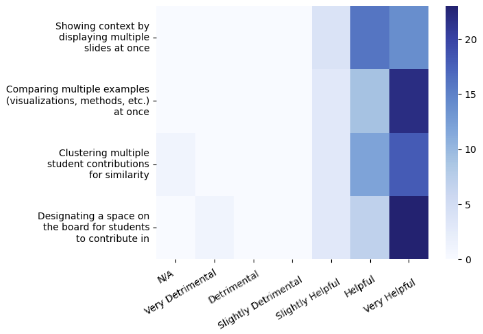}
    \caption{Student attitude towards design patterns used in class.}
    \label{fig:chart2}
\end{figure}

The attitudinal questions mixed statements with negative and positive sentiment.
Overall, results were positive. 
Looking at "I prefer to not use SAGE3," the 10 (20.4\%) students who agreed had issues like \emph{"The biggest issue is lag/latency and speed of actions."} and \emph{"make it faster? idk might just be my computer"} regarding technical performance, or comments such as \emph{"Can't zoom while cursor is over any type of content"} which showed some features went undiscovered by some students.

Comments from the 14 (28.5\%) students that disagreed with the statement "I prefer to not use SAGE3" included:
\begin{itemize}
    \item \emph{"good for collaboration, easy to have things visually present for ease of access"}
    \item \emph{"I like how it's really good for brainstorming"}
    \item \emph{"A much bigger screen and no switching between applications"}
    \item \emph{"It is more interactive compared to just simple powerpoint presentations. You can download material for the class directly from the board"}
    \item \emph{"SAGE3 made my learning experience easier for my teammates to collaborate. We also get to have fun and share different funny memes."}
\end{itemize}

The students that did not want to use SAGE3 used 14.2 of the 25 features we surveyed them on average, while those that did want to use SAGE3 of 16.7 of its features on average. 
We don't know if this difference shows correlation or causation. 
The features that seem to be of particular difference between the groups are: 
navigation, switch pages in a PDF, use annotations, use a webview to share a url, download files from the board, and use the wall outline for content placement.

\subsection{Evaluation 2}
In Fall 2023 and Spring 2024, we surveyed students in an Information Visualization graduate class that used SAGE3 and a large display for in-class instruction and interviewed the instructor in Spring 2024 about their experiences.
The class had collaborative projects, interactive discussions, and exercises designed to aid understanding of foundational information visualization concepts.
It was conducted in-person on the Virginia Tech campus by the same instructor from the SAGE group both semesters.
While SAGE3 was the substrate for delivering lectures and student activities, the university wide learning management system, Canvas, was also used for content delivery and assignment submissions.

A new survey for evaluating and improving SAGE3 was brought up in class near the end of Fall 2023 and in the latter half of Spring 2024, with not submitting it or expressing negative views on SAGE3 having no repercussions.
20 students from Fall 2023 and 14 students from Spring 2024 submitted the survey, making a total of 34 respondents.

The survey had three sections: 
the first briefly gauged how much they used SAGE3, the second prompted attitudinal responses on the use of SAGE3 with a big display, and the third focused on perceptions of the usability of SAGE3.
Following the Likert scale questions of the survey, students could write comments elaborating on their views.


\subsubsection{Survey Results}
\begin{figure}
    \centering
    \includegraphics[width=.85\columnwidth]{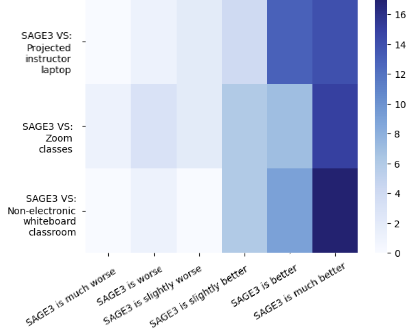}
    \caption{SAGE3 compared to other modes of teaching.}
    \label{fig:chart1}
\end{figure}

The survey responses for the first section on how frequently students used SAGE3 on their personal computer/laptop during class and outside class, are summarized below. The responses for the second section are summarized in Figures \ref{fig:chart1},\ref{fig:chart2}, and \ref{fig:chart3}, and the third section's responses also are summarized in Figure \ref{fig:chart3}.

All students used SAGE3 during class, but 6 (17.6\%) never used it outside class. 
4 of these 6 were fairly positive towards SAGE3; of these four, one felt an instructor projecting their laptop screen made for a better class experience than SAGE3, perhaps because they found using SAGE3 difficult based on their responses to the usability questions.
The 2 other students thought having lots of content on SAGE3 was distracting and felt overwhelmed by the amount of content; of these two, one felt Zoom classes were a better medium than SAGE3 while the other liked SAGE3 classes more than the three alternatives given.
These 2 students also found SAGE3 difficult to use and navigate in.

The attitudinal questions began with a list of statements on potential benefits of SAGE3 with a large display.
As seen in Figure \ref{fig:chart3}, students tended to agree this setup could benefit students on engagement, focus, ability to participate and contribute, enjoyment, and learning.


The next part of the attitudinal questions gave statements, some positive and others negative.
As seen in Figure \ref{fig:chart3}, the overall sentiment was positive, although a decent percentage of respondents at least slightly agreed that having lots of content on SAGE3 boards led to distraction (44.1\%) or that they felt overwhelmed by the amount of content on a board with the large display (23.5\%).


After the statement parts of the attitudinal questions, respondents rated 4 usage patterns for SAGE3 with the large display.
The patterns mentioned in this survey were not articulated as the more general patterns identified in this paper, although there is some relation between the questions and some of the patterns.
Specifically, the first 2 questions validate the Side-by-Side pattern, while the latter 2 validate the Spatial Rearrangement and Bespoke Spaces patterns.
As seen in Figure \ref{fig:chart2}, almost every respondent rated each usage pattern as at least slightly helpful.
We note that one respondent did not have a positive attitude to the patterns of dedicating space for students and clustering student contributions, explaining in the survey that \textit{"people can easily interrupt others' work. my work got deleted mistakenly by others. Sage3 is also not very fast and it takes time for it to update. (I would type something and it didn't show it until after a couple of minutes.)"}.


The final part of the attitudinal questions asked respondents to compare SAGE3 on the big display with 3 other instructions mediums, as seen in Figure \ref{fig:chart1}.
7 respondents in total (20.6\%) rated SAGE3 as at least slightly worse than at least one of the three listed instructions mediums, with one of these respondents preferring all three of the listed instruction mediums over SAGE3;
this respondent did not explain their ratings, unfortunately.
This 1 respondent took the class in Fall 2023, before certain improvements to SAGE3 were made.


Next came the usability section, whererespondents rated their agreement with a positive statement ("I found the SAGE3 user interface to be easy to use") and a negative one ("I found navigating within SAGE3 difficult");
the results of this are summarized in Figure \ref{fig:chart3}.
Overall, most students at least slightly agreed that the SAGE3 interface was easy to use, and a slight majority (52.9\%) felt navigation was difficult.




\subsubsection{Interview Results}
We interviewed the Information Visualization class instructor on the following topics:
comparing SAGE3 with a large display to prior teaching methods without SAGE3 or a large display, methods and activities for making the most of a large display with SAGE3, and benefits and challenges of using SAGE3 with a large display for visualization education.
We summarize interview findings not already covered below.

With SAGE3 and a large display, the instructor could show multiple 
slides, visualization examples, and more at the same time.
Without such an environment, they tended towards slide decks and having many web browser tabs open, which the instructor noticed had 2 key disadvantages when compared to SAGE3 with a large display: the traditional instruction style required more virtual navigation and limited collaboration and interactivity. Specific observations include:
    \begin{itemize}
        \item You can only show one item (slide, browser tab, etc.) at a time, limiting the ability to do tasks such as comparative analyses.
        \item You waste time scrolling, changing tabs, and doing other virtual navigation methods that could be faster with physical navigation.
        \item While doing virtual navigation (e.g. changing slides, tabs), you may forget part of what you were looking for, especially if you had to try several different tabs or slides to find the right one.
        \item Interactive exercises may have the instructor try to show students' ideas instead of enabling students to show what they are thinking.
        \item While some online boards like Miro or Google Jamboards give students space to sketch, such tools are limited by the smaller screen space available without a large display; this leads to virtual navigation being necessary, with all the problems that can bring.
    \end{itemize}

Given the expansive space a large display has and the affordances of content-rich canvases like SAGE3, our interview also covered best practices in such an environment.
Some examples the instructor gave are detailed in the section on Patterns above. 
The instructor preferred methods that inspired collaboration via content contribution and content rearrangement and used the space not only for semantic reasons but for class administration as well, such as tracking student participation by having them post something relevant on the board in a bespoke space.

Finally, we wrapped up with a discussion of the benefits of using SAGE3 with a large display for visualization education;
the instructor noted improved student engagement and higher-level learning.
They noticed students seemed much more engaged with the exercises during class and even expressed excitement after class.
In previous semesters, the instructor felt they were "pulling teeth" to get students to engage;
with SAGE3 and the large display, students were far more willing to talk about their ideas, contributions, and more.
Furthermore, the instructor noted that the ability to collaboratively explore many example designs and wrestle with relevant theoretical questions and ideas may help enable higher-level learning and analysis of the semantics of a visualization design space.
While this difference is the instructor's subjective experience, it does suggest the change in environment can provide, in their words, a "force multiplier" for learning.

The instructor did note one trade-off of using SAGE3 with a large display: when many items are on a board, it becomes laggy and slow. 
Thus, the instructor would create a new board about every two weeks.

\section{Discussion and Future Work}
Every person in our group uses a different mixture of these patterns in their classes, and this is influenced by both the instructor's preferences and the physical setups available in their classrooms. 
One group member has two large displays in their classroom, one at the front and one to the side. 
This layout makes it easy to use the \textit{Bespoke spaces} pattern, with the side board relegated to "class status" content including the class syllabus, presentation schedule, and assignment due dates. 
The main display is used in a \textit{Side-by-Side} manner to show lecture notes (given by this instructor as a web page) and example videos. 
Another team member relies primarily on the \textit{Content in Advance} pattern to arrange PDF files, video files, and links strategically on the board. 
This instructor also uses \textit{Burst of Content} and \textit{Spatial Rearrangement}, asking groups of students to iterate over their designs based on new material discussed in class.
During presentations, this instructor uses the \textit{Side-by-Side} approach with the presentation schedule in a webview on one side of the display and the presenting group screensharing on the other side. 
One member of our group with a large display that supports touch interaction uses annotations regularly to sketch charts. 

Even using the same tool, the variability in classroom environments and instruction methods leads to different teaching experiences, yet the patterns described above are general enough to be incorporated whenever a content-rich canvas and large display combination is available. 
The unifying factor is the "space to teach" afforded by a content-rich canvas like SAGE3 and large displays.
Our experience shows that our setups help us successfully engage students in learning required materials in ways that are collaborative and interactive, making headway in answering questions posed by Bach et al. \cite{10310184} regarding methods and environments for visualization and visually-intensive education.

A review of students' feedback does suggest weaknesses in our approach.
Content-rich canvases, such as SAGE3 and Miro, have a learning curve \cite{handley2023best} and many find navigation difficult. 
These difficulties could potentially have affected student motivation to learn a new system, which may have in turn affected, in our case, SAGE3 features explored and perceptions of usability like ease of navigation.
Students who didn't use SAGE3 often tended to not like it; 
this may be due to the aforementioned problems leading to low motivation to learn how to use it. 
We intend to increase efforts to train students to use SAGE3, evaluate and improve SAGE3's design and features for usability issues, and study this confounding factor in future evaluations.

Some students encountered technical problems, such as lag, which dampened their experience. 
This is a problem we are aware of; 
lag worsens if the board is too dense with content, and it is often better to branch into new boards when first experiencing delay. 
We should also note that SAGE3 is still under development and the software is improving all the time; 
many of the technical problems that were experienced in the Fall 2023 semester are already dealt with.
For more information about SAGE3 and its development, check out the SAGE3 Github \cite{sage3github} and wiki \cite{sage3wiki}.

The vast majority of students felt using a content-rich canvas like SAGE3 with large displays provided advantages over other forms of instructions: 
they appreciated that more could be shown on the large display, the support for group work, ability to download material, and how it left room for playfulness.

The biggest question raised as a result of the student evaluations is
\textbf{How much content is too much?} The surveys indicate that some students find the SAGE boards chaotic, distracting, even overwhelming. 
This perception suggests that there are factors in play with a large display plus content-rich canvas environment that could detract from the benefits such an environment can bring to educational settings.
Finding a good balance of content and techniques that utilize the collaborative, interactive affordances of large displays and content-rich canvases, like SAGE3, while also not overwhelming students with too much content remains an interesting avenue for future research.
Some example questions that could be answered along these lines include "how much content is too much", "what causes students to feel overwhelmed in a large display plus content-rich canvas environment", and "how can we address and mitigate the chaos while maintaining the benefits of large displays plus content-rich canvases?"

In the future we also anticipate that the SAGECells will have a bigger role in visualization classes since they can provide computational capabilities similar to a standard computational notebook but with the advantage of using a spatial arrangement, a feature that shows promise according to some studies \cite{harden2022exploring,harden2023there,harden2023exploring}.
In addition, developers have been creating plugins for SAGE3 and experimenting with running AI on SAGE3's backend. 
This opens possibilities that can affect visualization education, and merits future research. 

Perhaps one interesting outcome from using a content-rich canvas like SAGE3 for education is that it creates a visual representation, a visualization, of the lessons. 
Viewing a board after class conveys a sense for class progress and activities in a way that cannot be replicated in traditional classes.

\section{Conclusion}

This paper touches on challenges instructors come across in higher education, such as the need to utilize collaboration, interactivity, and play while teaching visually-intensive courses like visualization, especially with novel environments and tools that may be conducive for such courses. 
We bring up the concept of content-rich canvases, which are virtual surfaces containing many types of content meaningfully arranged spatially, and suggest using content-rich canvases in combination with large displays is beneficial for visually intensive courses. 

The heart of the paper details 6 usage patterns for an environment with at least 1 large display and a content-rich canvas software that we identified by analysing our own notes from teaching classes using the content-rich canvas software SAGE3 with large displays, which are Burst of Content, Side-by-Side, Content in Advance, Virtual Board Scroll, Bespoke Spaces, and Spatial Rearrangement. 
We discuss likely elements of the patterns in terms of layout, time, and lead actor. We also provide concrete examples for activities and their resulting boards taken directly from a visualization course.

We also incorporated evaluations that share students' attitude toward classes that use content-rich canvases, specifically in the form of SAGE3 boards. 
The students' impressions are vastly positive, feeling it improves collaboration, interactivity, and engagement.
However, they also expose some issues. 
Students that do not gain familiarity with the system seem to not be fond of content-rich canvases like SAGE3, indicating a need for more training as the semester starts as well as further study of what affects the difficulty of using a content-rich canvas in a collaborative setting and how such issues can be addressed through refined designs. 
Some technical issues like lag occur and can be a detriment; 
this can be solved by starting new boards when old boards become overburdened.
Most importantly, some students found the content-rich canvas approach, as done with SAGE3, overwhelming, which leads us to ask questions like "how much content is too much", a question we need to explore in future work.

 \acknowledgments{%
 	This project is funded in part by the National Science Foundation awards: 2004014, 2003800, 2003387, 2149133, and the Academy for Creative Media System.

 }

\bibliographystyle{abbrv-doi-hyperref}
\bibliography{casestudy}
\end{document}